\documentclass{aastex}

\newcommand{\ie}{{i.e.,} }
\newcommand{\eg}{{e.g.,} }

\newcommand{\etal}{{et al.}}

\newcommand{\ho}{H_0}                 
\newcommand{\qo}{q_0}                 
\newcommand{\B}{{$\blacktriangle$}}   
\newcommand{\pB}{{$\vartriangle$}}    

\begin{document}

\title{A CCD Study of the Environment of Seyfert Galaxies:}
\title{III. Host Galaxies and the Nearby Environments}

\author{S.N.~Virani\altaffilmark{1} and M.M.~De Robertis}
\affil{Dept.~of Physics and Astronomy, York University}
\affil{4700 Keele St., Toronto, ON, M3J 1P3, Canada}
\email{mmdr@yorku.ca, svirani@cfa.harvard.edu}

\author{M.L.~VanDalfsen}
\affil{Dept.~of Physics and Astronomy, McMaster University}
\affil{1280 Main St.~W., Hamilton, ON, L8S 4M1, Canada}
\email{vandalfs@physics.mcmaster.ca}

\altaffiltext{1}{Currently at Harvard-Smithsonian Center for Astrophysics,
60 Garden Street, MS 70 Cambridge, MA 02138}

\newpage

\begin{abstract}

A technique is described that permits the robust decomposition of the bulge
and disk components of a sample of Seyfert galaxies, as well as a (control)
sample of non-active galaxies matched to the Seyferts in the distributions
of redshift, luminosity and morphological classification.  The structural
parameters of the host galaxies in both samples are measured.  No
statistically significant differences at greater than the 95\% level are
found in these parameters according to a K-S test.

``Companion galaxies'' --- defined as any galaxy within a projected
separation of 200~h$^{-1}$ kpc from the center of the host --- are
identified and their basic properties measured.  A comparison between the
active and control samples in the distributions of apparent $R$ magnitude,
absolute $R$ magnitude (assuming the ``companions'' are at the distance of
the host), projected separation from the host, position angle relative to
the host, magnitude difference between the companion and host, and strength
of the tidal parameter, show no statistically significant differences.

Similarly, no statistically significant differences are found between the
control and active sample host galaxies in terms of light asymmetries ---
bars, rings, isophotal twisting, etc.  The implications for a model in
which interactions and mergers are responsible for inciting activity in
galactic nuclei are discussed briefly.

\vspace{0.7 truein}
\noindent
Accepted by the {\it Astronomical Journal}: 28 June 2000

\end{abstract}

\keywords{galaxies: active -- galaxies: evolution -- galaxies: interactions --
galaxies: Seyfert}

\newpage

\section{INTRODUCTION}

In order to understand the nature of activity in galactic nuclei, in
particular how it is initiated and maintained, it is reasonable to begin
with the question, how do active galaxies compare with non-active galaxies?
But any such statistical comparison is fraught with difficulty since this,
at least ideally, requires an understanding of the biases in the samples
selected.  Moreover, it is not clear just which parameter(s) should be
compared.  Could an active nucleus, especially in the more luminous members
of the active galactic nuclei (AGN) family, have a significant effect on
``macroscopic" scales; that is, not only on the immediate circumnuclear
environment, but even on the structural parameters of the host galaxy?

For these reasons, a comparison of the environments of AGN has been of
particular interest in the past two decades, motivated largely by the
widely held hypothesis that interactions and mergers play a significant
role in initiating activity in a nucleus.

Dultzin-Hacyan \etal\ (1999), De Robertis, Hayhoe \& Yee (1998a)
(hereafter Paper I), and Laurikainen \etal\ (1994) have observed that early
qualitative efforts to study the environments of AGNs were plagued by
sample biases and selection effects, especially in the selection of a
``control" sample, \ie a sample of non-active galaxies.

Seyfert galaxies have traditionally been the focus of such investigations
for the reasons articulated in Paper I: they are sufficiently nearby that
they can be studied in reasonable detail, and have a sufficiently high space
density to permit the compiling of reasonably large samples.  This is not
to say that the environments of both intrinsically higher luminosity AGN
such as QSOs, as well as lower luminosity systems such as LINERs and
``dwarf Seyfert galaxies" have not been investigated.  They have been, but
not nearly in such detail.

Recent studies of Seyfert galaxies which have taken more care, particularly
with the control sample such as Laurikainen \& Salo (1995), De Robertis,
Yee and Hayhoe (1998b) (hereafter, Paper II), and Dultzin-Hacyan \etal\
(1999) have achieved somewhat of a consensus compared with earlier studies:
Seyfert galaxies in general do not inhabit significantly richer environments
on scales $\ga 100$~kpc than do non-active galaxies.  Furthermore, it
appears that Seyfert 2 galaxies have an excess of companions, while Seyfert
1 galaxies inhabit environments which are somewhat deficient in companions.

On smaller scales and from a dynamical perspective, observations by Keel
(1996) and Kelm \etal\ (1998) indicate that there are no statistically
significant differences between groups and pairs of galaxies in which a
Seyfert is present, and those without an AGN.  (Interestingly, Kelm \etal\
find that Seyferts tend to avoid groups and close pairs with a low velocity
dispersion, as well as the closest pairs.)

Finally, on the smallest ``macroscopic'' scales, \ie on scales of the host
galaxies themselves, it is not clear whether there are significant
structural differences in galaxies hosting a Seyfert nucleus and
corresponding non-active galaxies.  Recent work by McLeod \& Rieke (1995),
Mulchaey \& Regan (1997), Mulchaey, Regan \& Kundu (1997), and Hunt \&
Malkan (1999) among others, have shown that Seyfert nuclei are not
located preferentially in kiloparsec-scale barred systems, though Hunt \& 
Malkan (1999) do suggest Seyferts may be associated with rings more often.
%
%

In Papers I and II, we explored the question raised at the outset by
quantitatively measuring the environments of both Seyferts and control
galaxies within a projected separation of $\approx 50-250$~kpc of the
active nucleus by computing the galaxy-galaxy covariance function
amplitudes.  In this work, we consider two other comparisons between the
active/non-active control samples: the characteristics of the ``companion
galaxies" within a projected distance of $200\, h^{-1}$~kpc (where
$\ho=50$~km~s$^{-1}$~Mpc$^{-1}$, $\qo=0.5$ and $h\equiv
\ho/50$~km~s$^{-1}$~Mpc$^{-1}$, hereafter) --- including faint companions
projected on the host itself and light asymmetries within the host --- as
well as the structural parameters of the host galaxies.

In \S 2 we describe the observations and techniques used to analyse the
data, while \S 3 presents the results.  The summary and conclusions are
contained in \S 4.

\section{OBSERVATIONS}

A study of this kind must necessarily be statistical in nature since the
redshift, and hence distance, of galaxies within a given angular radius is
unknown in general.  Foreground and background galaxy surface densities do
vary with position on the sky, but such differences should average out given
a sufficient number of fields distributed randomly on the sky, allowing
intrinsic differences in the distribution of actual or ``true'' companion
galaxies to become apparent.

A detailed description of the selection criteria for the active and control
samples is provided in Paper I.  Briefly, 32 Seyfert and 45 non-active
galaxies were culled from the CfA survey.\footnote{Papers I and II reported
on 34 Seyfert galaxies, while only 32 are presented herein.  This is
because two of the Seyfert galaxies in this sample are strongly
interacting, rendering the type of analysis presented in this paper highly
problematic.  As a result, these two systems were omitted from the
discussion.  This should not affect the basic conclusions.}
The distributions of redshift, morphological classification, as well as
luminosity of the non-active sample were chosen to match the Seyfert sample.
(Because a small, but non-negligible fraction of light in Seyfert galaxies
comes from the nucleus, the luminosity criterion for the non-active sample
was adjusted so that their mean absolute magnitude was roughly 10\% fainter
than the integrated Seyfert galaxy magnitude.)  The advantage of using a
CfA-selected sample (Huchra \etal\ 1983) is that it is believed to
be spectroscopically complete to an apparent Zwicky magnitude of 14.5,
permitting a reasonable matching of active and non-active galaxies and
avoiding a number of possible sample biases.  The disadvantage is that the
sample size is relatively limited.

Each of the galaxies was observed in photometric conditions through a
Cousins $R$ filter using the f/7.5 0.9~m telescope at KPNO over the period
1991 May $16-24$ UT.  A Tektronix $2048\times 2048$ CCD with $27\,\mu$m
pixels was used to record the data with a gain of 8.2 e$^-$~ADU$^{-1}$ and
read-noise of 13~e$^-$.  The focal-plane scale was
0\farcs{77}~pixel$^{-1}$.  The target galaxy was exposed typically for
900~s, centered approximately in the unvignetted field of
21\farcm{9}~$\times$~26\farcm{8}.  The data were taken under seeing
conditions of $1.4-3.0$~arcseconds.

The data were reduced using the Image Reduction and Analysis Facility
(IRAF)\footnote{IRAF is distributed by the National Optical Astronomy
Observatories, which are operated by the Association of Universities for
Research in Astronomy, Inc., under cooperative agreement with the National
Science Foundation.}  and calibrated by means of Landolt (1992) standard
fields in a straightforward manner.

It is important to note the advantages of CCD photometry compared with
previous photographic surveys: in the first place, photometric accuracy
is significantly better.  This is important, not only for internal
consistency among fields, but for quantitative comparisons with other
studies, especially QSOs (see, \eg Paper II).  Secondly, the nucleus and
disk are not saturated, permitting the search for companion systems and
light asymmetries throughout the image, not just well outside the disk.

\section{REDUCTIONS AND ANALYSIS}


\subsection{\it Surface Brightness Profiles and Structural Parameters}

The apparent magnitudes of the host galaxies were measured in circular
apertures with an appropriately sized sky-background annulus using the
Picture Processing Package (PPP) of Yee (1991).  The magnitudes typically
converged to a couple of hundredths of a magnitude.  Light from
foreground stars was subtracted from the galaxy's integrated magnitude if
the star was within the relevant aperture. The magnitudes were also
corrected for Galactic absorption (as provided by the NASA/IPAC
Extragalactic Database (NED)\footnote{The NASA/IPAC Extragalactic Database
(NED) is operated by the Jet Propulsion Laboratory, California Institute of
Technology, under contract with the National Aeronautics and Space
Administration.}, where $A_R = 0.55\,A_B$) when necessary.

One method of measuring the surface-brightness profiles of Seyfert galaxies
is via a three-component, chi-squared minimization procedure (\eg
Alonso-Herrero, Ward, \& Kotilainen 1996; and Kotilainen, Ward, \& Williger
1993). That is, one could model the surface-brightness profile of a Seyfert
galaxy by simultaneously modeling the nuclear, bulge, and disk components
using appropriate fitting functions.  This technique can be problematic,
however, for data with a finite signal-to-noise (S/N) ratio, with relatively
strong nuclear contamination of the bulge, leading to numerical
instabilities and concerns about uniqueness.

One simplification is to ignore data within two seeing disks within the center
of the galaxy (normally 5 pixels), and to then restrict the algorithm to a
two-component fit---bulge plus disk---from which a nuclear 
contribution may be inferred.

Experiments employing artificial data showed that while the disk parameters
could be well recovered using either method, the bulge parameters showed a
considerable degree of variability in some galaxies. Thus, both models have
limited utility in this context.

A third technique adopts an empirical approach for modeling
the nuclear component. Since Seyfert nuclei are point sources when they
have distinct nuclei, (eg., Nelson \etal\ 1996; Malkan, Gorjian, \& Tam
1998), it is reasonable to account for this component by subtracting a
shifted profile of a scaled, high S/N ratio, unsaturated and unblended star
near the host nucleus (\ie a good point-spread function, PSF), using a
scale factor based upon pre-determined fitting criteria.
McLeod and Rieke (1995) used this latter technique in their analysis.  They
assumed the nuclear contribution to be that of a point source and then
proceeded to subtract larger fractions of a PSF until the resulting profile
started to turn over.  We refer to this as the {\em shift, scale, and
subtract technique}, or ``SSS-technique.''  The resulting light
distribution should then be virtually pure bulge $+$ disk light and, hence,
amenable to a 2-component fit. To be sure, this approach removes too much
of the nuclear light, as artificial tests reveal, but so long as the fit
ignores the inner 2 seeing disks, a reasonable two-component fit is
produced for typical nuclear contributions.  McLeod and Rieke (1995)
obtained the same result; one achieves robust fits for a wide variety 
of nuclear brightnesses.
The actual nuclear component is then found by subtracting the bulge and
disk light calculated from the fitted structural parameters from the total
light in the galaxy within some fiducial radius.

The IRAF/STSDAS Ellipse task was then used to generate the azimuthally
averaged surface brightness profile with the nuclear component removed;
that is, the intensity as a function of the semi-major axis (as well as
ellipticity, position angle, ellipse center and ellipse harmonics) were
computed along with the associated formal uncertainties.

The resulting surface-brightness profile was then modeled to determine the
host-galaxy's bulge and disk structural parameters using a two-component
fit (see below) consisting of an $r^{1/4}$ bulge (de Vaucouleurs 1948) and
an exponential disk (Freeman 1970).  Once the two structural parameters for
each component were determined, a Gaussian PSF with the appropriate FWHM
based on the seeing in the image was used to model the nucleus itself.

In linear units, it will be assumed that the intensity $I$ of a Seyfert
galaxy's profile at a radius $r$ can be expressed:
\begin{equation}\label{profileeqn}
I(r) = I_e\,e^{-7.688\left[\left(r/r_e\right)^{1/4} -1\right]} + 
I_d\,e^{-r/r_o} + I_n\,e^{-r^2/2\sigma^2}
\end{equation}
where the factors $I_e$ and $r_e$ are the intensity (at $r_e$) and scale
radius for the (de Vaucouleurs) bulge, $I_d$ and $r_o$ are the central
intensity and scale radius for the (exponential) disk, while $I_n$ and
$\sigma$ are the central intensity and standard deviation of the Gaussian
PSF.

There are alternative functional forms for the bulge, but while a
de Vaucouleurs law is not unique in its ability to fit the profiles of
elliptical galaxies or bulges, the actual differences between it and other
empirical laws are relatively small.

Virani \& De Robertis (2000) utilize the structural parameters determined
by fitting the surface brightness profile for each galaxy, as well as the
nuclear component data, to search for intrinsic differences between
Seyfert 1s and Seyfert 2s. The focus of this paper, however, is simply to
compare the entire Seyfert population to the set of control galaxies.

The data were fit to Equation~(1) using a robust non-linear least squares
routine (Press \etal\ 1992). In this algorithm, the relative intensities
for the bulge and disk, $I_e$ and $I_d$, as well as the scale radii, $r_e$
and $r_o$, are determined along with their formal uncertainties once the
function's chi-squared achieves a global minimum.  The reduced chi-squared
is also reported.  (Note that $\sigma$ in Equation~(1) is held constant.)
The routine was tested at some length using artificial data sets
representative of a galaxy with a pure bulge, with a pure disk, with a bulge
and disk in varying proportions, and with the addition of a nuclear
component of various strengths.  In all cases we found good agreement
between the code's results and the input data parameters.  Moreover, the
routine was designed so that data from the innermost pixels (roughly
equivalent to two seeing disks) were not used.  This provided us with the
additional assurance that we were fitting a bulge plus disk light profile
and further mitigated the small contribution from any remaining nuclear
component. The efficacy and accuracy of our computer algorithm is more
thoroughly explored and discussed in Virani \& De Robertis (2000).

The results of this analysis are recorded in Tables $1-6$.  Table 1 lists
the structural parameters with their formal uncertainty for the control
sample; $r_o$ and $r_e$ are the scale radii for the disk and bulge
respectively, while $\mu_d$ and $\mu_b$ are the appropriate surface
brightnesses corresponding to $I_d$ and $I_e$.
Table 2 shows the fraction of light in the bulge and disk for each system,
the bulge-to-disk ratio (B/D), as well as the total (integrated) apparent
$R$ magnitude for each galaxy in the control sample.  Tables 3 and 4 give
the structural parameters for the Seyfert 1s and 2s respectively, while
Tables 5 and 6 show the fraction of light in each of the three components:
disk, bulge and nuclear, as well as the their bulge-to-disk ratio and
integrated $R$ magnitude.

Figure 1 shows the distribution of the disk scale radius (top) and bulge
scale radius for radii less than 7 kpc. (While there are scale radii larger
then 7 kpc, these also contain proportionally very large uncertainties.)
The solid line shows the control sample, while the dashed line is for the
Seyfert galaxies (types 1 and 2s together).  Figure 2 shows the
distribution of disk (top) and bulge (bottom) surface brightnesses.  Figure
3 illustrates the distribution of the bulge-to-disk ratios.
Kolmogorov-Smirnov (K-S) tests show that there are no significant
differences between the control and active samples in these structural
parameters.


\subsection{\it Companion Galaxy Characteristics}

Unlike Paper I in which object classification and analysis were performed
entirely using PPP, all non-stellar objects within a projected radius of
$200\, h^{-1}$~kpc were identified using both visual inspection and the
shape of the light profile compared with a local stellar profile.  There
was excellent agreement between the sample of companion galaxies detected
in an automated fashion with PPP with those detected manually.  By adding
artificial companion galaxies to a variety of real Seyfert host galaxies,
it was determined that typically galaxies brighter than $R \approx +18.5$
(corresponding to an absolute magnitude of $M_R \approx -16.5$, slightly
fainter than the SMC at the mean redshift of the sample) could be recovered
outside the immediate nucleus.

A search for companion galaxies and light asymmetries projected on the disk
of a host galaxy was performed using unsharp masking techniques.  That is,
the data were convolved using a flux-preserving Gaussian kernel with a
full-width at half-maximum (FWHM) equal to four times the seeing.  The
original image was then divided by the convolved image to yield an
``unsharp-masked'' image.  Light asymmetries are more readily identified in
the masked image and can be classified in a qualitative sense as either
spiral arms, bars, rings, isophotal twisting and/or very close companion
galaxy.

The instrumental magnitude and position of each companion galaxy relative
to the nucleus of the host galaxy were measured.  These were converted to
an apparent magnitude and projected separation using the photometric
calibration and redshift of the host galaxy under the assumption that the
companion lies at the same distance as the host.  Figure 4a provides a
histogram of the distribution of apparent magnitudes for the relevant
samples, while Figure 4b shows the corresponding absolute magnitude
distributions under the assumption that the companion galaxies are situated
at the same redshift as the host.

All histograms follow the same format: the solid line indicates the control
sample, while the dashed line shows the Seyfert sample (1s and 2s
combined).  The histograms have been normalized such that the area under each
curve is unity.  Quantitative comparisons between distributions using the
K-S test in which the {\it null hypothesis} is that the two data sets
(properties) being compared are from the same underlying population.
Differences are significant only when the K-S test rejects the null
hypothesis at a confidence level of 95\% or greater.  The results from the
K-S test on both distributions show that the null hypothesis cannot be
rejected, apart from a marginal agreement between the central bulge surface
brightnesses, $\mu_b$, of the control sample and Seyfert 2s.

Within a projected radius of 200~kpc, 359 optical companions were found
around the 32 Seyfert host galaxies for a companion frequency of
$11.2\pm1.0$ companions/host (175 around Seyfert 1 and 184 around Seyfert
2), and 520 optical companions were found around the 47 control host galaxies
for a companion frequency of $10.6\pm0.9$ companions/host (where the
uncertainties are the root-mean-square of the mean).  Figure 5 shows the
relative frequency of the number of companions around each host.  The K-S
test indicates that the distributions are similar for the two samples.

Figure 6 illustrates the distribution of projected separations, $s$,
between the centers of a host and companion, assuming they are situated at
the same distance.  While the K-S test suggests that both distributions are
similar, the Seyfert sample has a marginally higher companion frequency
within 50 kpc.  Figure 7 depicts the distribution of position angles of
companion galaxies with respect to the host galaxy.  The histogram is binned
into $45^\circ$ intervals.  The observed distributions are consistent with
a uniform random distribution as might be expected.  Figure 8 shows the
distribution of the logarithm of the magnitude difference between the host
and companion galaxies, assuming once again that the companions lie at the
same distance as the host.  Note that at $R \approx +19$ the distribution
turns over and rapidly drops to zero, indicating the limiting magnitude to
which companions could be detected efficiently.

The maximum tidal influence of a companion galaxy on its host galaxy,
$Q_i$, is proportional to the companion's mass divided by the cube of its
projected separation, $s_i$.  Under the assumption that light traces mass
and that the companions are situated at the redshift of the host, then the
maximum tidal parameter for each host $Q \propto \sum_{i=1}^{N} Q_i$ where
$Q_i \propto {L_i}/{s_i^3}$ and $L_i$ is the luminosity of the companion.
Figure 9 indicates the distribution of this parameter.  A K-S test shows
that the distribution of this tidal parameter is similar for both the
Seyfert and control samples.  Moreover, Seyfert 1s and 2s have similar
tidal parameter distributions.  There are a few hosts in both samples for
which $Q\!>\!10^5\,\,L^\ast/{\rm Mpc}^3$ implying severely disturbed
systems.  ($L^\ast$ here is the equivalent luminosity for an $M^\ast_R =
-22.1$ galaxy; Schechter 1976.)

As with all the parameters related to the optical companions of the host
galaxies, the environments of Seyfert galaxies and the control galaxies are
very similar.  There is a difference in the distribution of faint galaxies
around Seyfert 1s compared with Seyfert 2s, though when considered
together, the Seyfert and a non-active (control) companion distributions are
similar.  There is also no obvious difference in the tidal influences the
companion galaxies have on the hosts, though both samples have their share
of very tidally disturbed systems ($\sim\!8\,\%$).

\subsubsection{Host Galaxy Light Asymmetries}

Light asymmetries and morphological disturbances with host galaxies are
important features to consider since they {\it may} be a symptom of a
recent interaction or merger or evidence for a radial flow of material.
Though such features can only be characterized in a qualitative sense, it
might be expected that a sample of active galaxies would show a greater
departure from ``normalcy'' than a non-active sample; \ie from typical
exponential (spiral) disks and featureless bulges.

Using primarily the unsharp masking technique, each host galaxy was
searched for the following features: bars, rings, significant isophotal
twisting, and other features that could be evidence for a recent
interaction: \eg tidal tails, bridges, and (asymmetric) prominent dust
lanes.

Bars have long been considered an efficient mechanism for transporting gas
to the sub-kpc regions of a galaxy, and hence, an AGN (\eg Athanassoula 1992;
Shlosman \& Noguchi 1993).  Inner and outer rings are also formed during
``disturbances'' (\eg Hunt \& Malkan 1999) or when a companion galaxy
passes right through the host galaxy (\eg Combes \etal  1991).  Other
distortions may also provide circumstantial evidence for a recent
interaction such as tidal tails, bridges, prominent dust lanes, or other
significant light asymmetries.  Similarly, an extreme twisting of
isophotes, where the position angle well outside the nucleus changes by a
large amount, in this case $\pm 45\degr$, could also flag a previous
disturbance. 

Tables~7 and 8 show the frequency of such morphological disturbances in
both the Seyfert host and the control host respectively detected in this
analysis.  A (\B)\ indicates that the feature was noticed in the galaxy,
and (\pB)\ indicates that only a partial feature was observed. Column 1
gives the name of the galaxy, columns $2-5$ indicate a Bar, Ring,
Distortion, or large Position-Angle excursion for each system respectively,
while column 6 shows the occurrence of any of the previous disturbances
(i.e. bar and/or ring and/or distortions).

Table~9 summarizes the frequency of bars, rings, and other distortions.  It
presents the number of galaxies containing the feature (including partial
features in parentheses) as well as the percentage of galaxies with that
feature.  As can be seen, there are slightly different ratios of galaxies
that contain bars or rings in both samples. However, a large fraction of
galaxies in both samples contain some form of disturbance, although a
slightly larger fraction of Seyfert galaxies contain a disturbance.

Acknowledging the qualitative nature of this classification, there do not
appear to be large differences between the active and non-active samples.
Certainly there are no differences in the frequency of bars and
``distortions.''  Seyfert galaxies may have rings somewhat more frequently
and exhibit large position-angle excursions, but the significance is
marginal at best.  It appears that in terms of light asymmetries and
disturbances, Seyfert galaxy hosts are not significantly different from the
hosts of the non-active sample.

In a naive interpretation of the interaction or merger hypothesis for
active galaxies, differences within the local environments and perhaps even
host-galaxy properties might be expected between Seyfert galaxies and a
reasonably matched control sample.  At the level of investigation presented
in this paper, however, no statistically significant differences are
uncovered.  When combined with the analyses presented in Papers I and II,
it appears that if the interaction/merger model is correct, then its
interpretation must be more complex than imagined.

\section{CONCLUSION}

A technique was developed that allowed the measurement of the structural
parameters of the bulge and disk in the hosts in a sample of Seyfert and
control (non-active) galaxies.  No statistically significant differences
were found in these parameters according to the K-S test.

A comparison between the properties of the companion galaxies within
200~$h^{-1}$~kpc of each host in both the active and control samples ---
\ie the distributions of apparent $R$ magnitude, absolute $R$ magnitude
(assuming the ``companions'' are at the distance of the host), projected
separation from the host, position angle relative to the host, magnitude
difference between the companion and host, and strength of the tidal
parameter --- show no statistically significant differences.

Moreover, no statistically significant differences were found between the
control and active sample host galaxies in terms of light asymmetries ---
bars, rings, isophotal twisting, etc.  

It appears that the nearby environment of Seyfert galaxies is not
significantly different from non-active galaxies with the same
morphological distribution; that is, the companion characteristics are
indistinguishable, as are the structural parameters and a qualitative
assessment of any light asymmetries.

Differences in some of these properties might have been anticipated
according to a simple interpretation of the interaction or merger
hypothesis for the initiation of activity in galactic nuclei.  That none
was found may indicate that if this hypothesis is correct, it must be
operating on a more complex level.

\acknowledgements 

The analyses presented in this paper are largely the result of MSc theses
by VanDalfsen (1997) and Virani (2000).  M.M.D.R. gratefully acknowledges
financial support by the Natural Sciences and Engineering Research Council
of Canada on which this research was based.

\newpage


\newpage


\figcaption[fig1.ps]{Relative distribution of disk scale lengths (top) and
bulge scale lengths (bottom) for radii less than 7 kpc.  Control galaxies
are shown with a solid line, while Seyfert galaxies (1s and 2s) by a dashed
line hereafter.}

\figcaption[fig2.ps]{Relative distribution of disk central surface
brightnesses (top) and bulge central surface brightnesses (bottom).}

\figcaption[fig3.ps]{Relative distribution of bulge-to-disk ratios for
control galaxies (solid line) and Seyfert galaxies (dashed line) for
ratios less than 2.}

\figcaption[fig4.ps]{Relative distribution of $R$ apparent magnitudes (top)
and $R$ absolute magnitudes (bottom) of companion galaxies within a
projected radius of 200 kpc of the main galaxy.  In the latter case, the
companion galaxies are assumed to have the same distance as the main galaxy.}

\figcaption[fig5.ps]{Relative distribution of the number of companion
galaxies within 200 kpc of the main galaxy.}

\figcaption[fig6.ps]{Relative distribution of the projected separation of
companion galaxies from the main galaxy for distances less than 200 kpc.}

\figcaption[fig7.ps]{Relative distribution of the position angle of the
companion galaxies with respect to the main galaxy for distances less than
200 kpc.}

\figcaption[fig8.ps]{Relative distribution of the difference in $R$ apparent
magnitude between the companion galaxy and the host or main galaxy for
companions less than 200 kpc.}

\figcaption[fig9.ps]{Relative distribution of the tidal strength
of companion galaxies surrounding the host or main galaxy for companions
less than 200 kpc.}


\begin{references}

\reference{}Alonso-Herrero, A., Ward, M.J., \& Kotilainen, J.K. 1996,
\mnras, 278, 902

\reference{}Athanassoula, E. 1992, \mnras, 259, 345

\reference{}Combes, F., Boiss\'e, P., Mazure, A., \& Blanchard, A. 1991,
In Galaxies and Cosmology, Springer-Verlag.

\reference{}De Robertis, M.M., Hayhoe, K., \& Yee, H.K.C. 1998a, \apjs,
115, 163 (Paper I)

\reference{}De Robertis, M.M., Yee, H.K.C., \& Hayhoe, K. 1998b, \apj, 496,
93 (Paper II)

\reference{}de Vaucouleurs, G. 1948, Ann. d'Astrophys., 11, 247

\reference{}Dultzin-Hacyan, D., Krongold, Y., Fuentes-Guridi, I., \& Marziani, P. 1999a, \apjl, 513, 111

\reference{}Freeman, K. 1970, \apj, 160, 811

\reference{}Huchra, J., Davis, M., Latham, D.., \& Tonry, J. 1983, \apjs,
52, 82

\reference{}Hunt, L.K., \& Malkan, M.A. 1999, \apj, 516, 660

\reference{}Keel, W.C. 1996, \aj, 111, 696

\reference{}Kelm, B., Focardi, P., \& Palumbo, G.G.C. 1998, \aap, 335, 912

\reference{}Kotilainen, J., Ward, M.J., \& Williger, G. 1993, \mnras, 263, 655

\reference{}Landolt, A. 1992, \aj, 104, 340

\reference{}Laurikainen, E., \& Salo, H. 1995, \aap, 293, 683

\reference{}Laurikainen, E., Salo, H., Teerikorpi, P., \& Petrov, G. 1994, \aaps, 108, 491

\reference{}Malkan, M.A., Gorjian, V., \& Tam, R. 1998, \apjs, 117, 25

\reference{}McLeod, K., \& Rieke, G. 1995, \apj, 441, 96

\reference{}Mulchaey, J.S., \& Regan, M.W. 1997, \apjl, 482, 135

\reference{}Mulchaey, J.S., Regan, M.W., \& Kundu, A. 1997, \apjs, 110, 299

\reference{}Nelson, C.H., Mackenty, J.W., Simkin, S.M., \& Griffiths, R.E. 1996, \apj, 466, 713

\reference{}Press, W., Flannery, B., Teudolsky, S., \& Vetterling,
W. 1992. Numerical Recipes in FORTRAN: The Art of Scientific Computing (2
ed.) Cambridge University Press.


\reference{}Schechter, P. 1976, \apj, 203, 297

\reference{}Shlosman, I., \& Noguchi, M. 1993, \apj, 414, 474

\reference{}VanDalfsen, M. 1997.  MSc Thesis. The Nearby Environment of Seyfert Galaxies: A comparative Study

\reference{}Virani, S. 1999.  MSc Thesis. In preparation.

\reference{}Yee, H.K.C. 1991. \pasp, 103, 396

\end{references}
\end{document}